\documentstyle[preprint,prl,aps,epsf]{revtex}

\begin{document}

\title{Evidence for the absence of regularization corrections to the 
partial-wave renormalization procedure in one-loop self energy calculations in 
external fields}
\author{Igor Goidenko$^{1,2}$, G\"unter Plunien$^{2}$, 
Sven Zschocke$^{3}$, Leonti Labzowsky$^{1,4}$, Gerhard Soff$^{2}$}
\address{$^{1}$Institute of Physics, St.Petersburg State University,
198904 Uljanovskaya 1, Petrodvorets, St.Petersburg, Russia}
\address{$^{2}$Institut f\"ur Theoretische Physik, Technische  Universit\"at
Dresden, Mommsenstrasse 13, D-01062  Dresden, Germany}
\address{$^{3}$Forschungszentrum Rossendorf, PF 510119, 01314 Dresden, Germany}
\address{$^{4}$Max-Planck-Institut f\"ur Physik komplexer Systeme, 
N\"othnitzer Strasse 38, D-01187 Dresden, Germany}
\date{\today}
\maketitle

\begin{abstract}
The equivalence of the covariant renormalization and the partial-wave renormalization (PWR) 
approach is proven explicitly for  the one-loop self-energy correction (SE) of a bound electron 
state in the presence of external perturbation potentials. No spurious correction terms
to the noncovariant PWR scheme are generated for Coulomb-type screening potentials and
for external magnetic fields.
It is shown that in numerical calculations of the SE with Coulombic perturbation
potential spurious terms result from an improper treatment of the unphysical
high-energy contribution.
A method for performing the PWR utilizing the relativistic B-spline approach
for the construction of the Dirac spectrum in external magnetic fields
is proposed. This method is applied for calculating
QED corrections to the bound-electron $g$-factor in H-like ions.
Within the level of accuracy of about 0.1\% no spurious terms are generated in
numerical calculations of the SE in magnetic fields.


\end{abstract}

\section{Introduction}
The partial-wave renormalization (PWR) approach has been proposed a few   
years ago \cite{pls93,qag93} as a convenient but 
non-covariant method to perform the 
renormalization numerically in bound-state QED calculations. 
It has been successfully applied
first in exact numerical calculations of the self-energy 
and vacuum-polarization correction of order $\alpha$ 
\cite{pls93,lea93a,pea93} 
and further applied in exact calculations of QED-corrections of order 
$\alpha^2$ \cite{lea93b,pea96} ($\alpha$ is the fine-structure constant).
A fair agreement between the results obtained within
different numerical approaches can be stated (see, e.g., the
results for the effective self-energy correction in Refs. \cite{pea96,mas96}).
Nevertheless, questions about the equivalence between 
the covariant renormalization
and the numerical PWR scheme respectively conjectures about the possible
occurrence of {\em spurious terms} in numerical calculations of higher-order
QED effects have been raised in the past \cite{pss98,yas97}. 
In  Ref. \cite{yas97} this issue has been anticipated qualitatively 
in connection with problems encountered in the numerical
evaluation of the screened Lamb shift
when non-covariant, numerical renormalization schemes are employed.
Persson {\em et al.} \cite{pss98} have made the first attempt to derive such
spurious correction terms to the PWR analytically. 
To our knowledge this has been the first and only reference 
in which corresponding terms have been presented explicitly.
They considered the exact self-energy correction of a bound
electron state in the presence of an additional Coulomb-type screening
potential $V_c$
which is treated perturbatively. Formulating the PWR by employing the
Pauli-Villars regularization a generic, regulator-independent correction
term that could contribute to the level shift of a bound state $|a\rangle$ is
derived from corresponding counterterms (see Eq. (44) of Ref. \cite{pss98}):
\begin{eqnarray}
{\cal E}^a (\Lambda\rightarrow \infty) &=& - \frac{\alpha}{2\pi}\,
\langle a|(V_c(r) -\overline{V}_c)\,\ln (r)|a\rangle\quad ,
\overline{V}_c = \langle a| V_c (r) | a \rangle .
\label{eqin}
\end{eqnarray}
Although for the particular situation under consideration the correction
term cancels because it occurs with opposite sign 
in different subgroups of diagrams, 
the authors conjectured that this may not always be the case in 
calculations of higher-order QED effects and that the PWR and the covariant
renormalization could lead to results that differ generically by terms
of the form of Eq. (\ref{eqin}).
The conclusions drawn in \cite{pss98} received further the support from numerical
results for the SE including a perturbing $1/r$ potential \cite{Yer}.

In this paper we wish to address first the question about the occurrence of 
spurious terms of the generic type (\ref{eqin}). Therefore, we reinvestigate
the problem considered in Ref. \cite{pss98}. 
In conclusion we find no indication for
spurious terms generated by the numerical PWR method in 
the case of one-loop SE calculations in 
external fields. The correction (1) is shown to be due to
an improper treatment of the unphysical high-energy contribution to the SE.

Recalculating
the examples, given in Ref.\cite{Yer} we find that the spurious terms 
originate again from
a similar unphysical high-energy contribution.

Finally, we investigate the problem of the spurious terms in 
an external magnetic field.
In Ref.\cite{pss98} it was conjectured that in the case of external magnetic perturbation
spurious contribution to SE should remain. 
Employing a new approach developed here
for the PWR in magnetic field, which 
is based on basis set expansion for the Dirac
equation due to the Chen and Goldman \cite{gold} we 
prove the absence of the spurious terms for the QED
corrections to the bound electron $g$-factor 
numerically on the level of accuracy of about 0.1\%.

\section{Equivalence between the covariant and the partial-wave renormalization}
In Ref. \cite{pss98} a generic correction term between the PWR and the covariant
renormalization contributing to the energy shift of a bound-electron state 
$|a\rangle$ interacting with a spherically symmetric perturbation potential
$V_c$ has been derived within the Pauli-Villars regularization scheme.
The authors obtain 
a spurious correction term to the PWR from both the $\Lambda$-dependent
wave-function correction and from the vertex correction 
(see Eqs. (36) and (39) of Ref. \cite{pss98}): 
\begin{eqnarray}
{\cal E}^a (\Lambda) = 
\frac{2\alpha}{\pi}\,\sum_{\ell = 0}^{\infty}\,\int_0^\infty \,dk\,
\frac{k^2\,(2\ell + 1)}{k'\,(k+k')^2}\,
\langle a| j^2_\ell (kr) (V_c(r)-\overline{V}_c)|a\rangle  
= \frac{2\alpha}{\pi}\,\sum_{\ell = 0}^{\infty}\,\int_0^\infty \,dk\,
f_\ell^a (k,\Lambda)\quad , 
\label{eq1}
\end{eqnarray}
where $\Lambda$ denotes the Pauli-Villars regulator and 
$k' = \sqrt{k^2+\Lambda^2}$. 
Without going through the details of the derivations given in 
Ref. \cite{pss98} we take Eq. (\ref{eq1})
as the starting point of the following considerations.
Note that the integral over the momentum $k$ will be finite for each 
partial wave $\ell$. The $k$- and $r$-dependence of the integrand ensures
a sufficient convergence of both integrals, which allows one to interchange
the order of integrations. 
For any finite value $r > 0$ the integral over $k$ is 
sufficiently convergent, i.e., the integrand falls off as $\sim 1/k^3$ 
for asymptotic values of $k$. 
On the other hand the contribution to the matrix element
arising from the integration over $r$ from $r=0$ to some arbitrarily small
value $r=r_0$ will be negligible. In the following considerations 
(part A) we keep the usual order of integrations as dictated by the
PWR approach (see e.g. \cite{pls93}), i.e. that the matrix element should 
be evaluated before the integration over $k$ is performed.

Suppose we could interchange the summation over $\ell$ with
all the integrations involved, then the correction term 
Eq. (\ref{eq1}) vanishes evidently in view of the identity
\begin{eqnarray}
\sum_{\ell = 0}^{\infty}\,(2\ell + 1) j^2_\ell (kr) = 1
\label{eq2}
\end{eqnarray}
after the integration over $r$ is performed.
Accordingly, the PWR approach and the covariant renormalization would be  
equivalent leading to identical results for the renormalized energy shift.
Thus, one could try to prove explicitly whether or not Weierstrass' theorem
for uniformly convergent functional series holds in case of the
generic correction term (\ref{eq1}). 
In the following we shall demonstrate that the functional series 
(\ref{eq1}) is uniformly convergent.

\subsection{Proof of uniform convergence}
The infinite summation over partial waves $\ell$ may be decomposed into
a finite sum $0\leq \ell \leq L-1$, with $L \gg 1$ and the remaining
infinite sum over $\ell\geq L$. Accordingly, it is sufficient to focus 
on the remaining infinite sum. 
For this purpose we may substitute $k=t/(1-t)$, which transforms the 
indefinite integral over $k$ involved in 
the generic expression (\ref{eq1}) into a definite integral: 
\begin{eqnarray}
{\cal E}_L^a (\Lambda) 
&=& \frac{2\alpha}{\pi}\,\sum_{\ell = L}^{\infty}\,\int_0^1\, dt\,
f_\ell^a \left(t/(1-t),\Lambda\right) \quad .
\label{eq3}
\end{eqnarray}
As the next step we have to find an upper bound (majorante) $u^a_\ell$ 
for each term $f_\ell^a$ of the functional series for all $t\in [0,1]$ 
and for a fixed but large $\Lambda \gg 1$ such that 
$\left|f^a_\ell (t/(1-t),\Lambda)\right| 
< u^a_\ell (\Lambda)$ 
and $\sum_{\ell = L}^{\infty}\,u_\ell^a (\Lambda) = C^a(\Lambda) < \infty $
holds.

If Weierstrass' criterium of uniform convergence is valid, we can interchange
the summation over $\ell$ with the integration over $t$ and thus over $r$
as well. 
One key-point for finding a convergent majorante series is to 
estimate appropriately the square of the spherical Bessel functions involved.  
We rewrite Eq. (\ref{eq3}) in the following form:
\begin{eqnarray}
{\cal E}_L^a (\Lambda) = \frac{2\alpha}{\pi}\,\sum_{\ell = L}^{\infty}\,
\int_0^1\, dt\,\frac{t^{2+\beta}}{(1-t)^{1+\beta}}
\frac{(2\ell +1)\,\langle a|\left[
\left(\frac{1-t}{rt}\right)^\beta\,
j_\ell^2 (\frac{rt}{1-t})\right]\,r^\beta\, 
(V_c (r) - \overline{V}_c)|a\rangle}
{\sqrt{t^2+(1-t)^2\Lambda^2}\,\left[t+\sqrt{t^2+(1-t)^2\Lambda^2}\right]^2}
\quad ,
\label{eq6}
\end{eqnarray}
where  $\beta > 0$. 
Note, that the integrand of the integral (5) has a
complicated analytical structure when extended into 
the complex $t$-plane. Accordingly, the integral over $t$ must be
undestood as being performed along
a suitable chosen contour from the very beginning.
The choice of the contour of integration will be done below.  The integral (5) is
difficult to handle analytically in closed form. Therefore,
we may now employ the following estimates for the two factors
involved in the integrand:
\begin{eqnarray}
\left| \frac{}{}
\right\{\sqrt{t^2+(1-t)^2\Lambda^2}\,
\left[t+\sqrt{t^2+(1-t)^2\Lambda^2}\right]^2\left\}^{-1}
\frac{}{} \right| \leq
\frac{1}{4}\quad ,\quad \Lambda \gg 1\quad,
\label{eq7a}
\end{eqnarray}
and for all $\ell \geq L \gg 1$ and for all 
complex arguments $z$ inside of a bounded region of
the complex $t$-plane (see appendix A)
\begin{eqnarray}
\left| F_\ell (z,\beta) \right| &=& ß\left| \frac{j_\ell^2(z)}{z^\beta} \right| \leq
(1+\beta)\, \frac{j_\ell^2(a'_{\ell ,1})}{(a'_{\ell ,1})^\beta} \leq 
c_\beta\, \left(\ell + \frac{1}{2} \right)^{-\frac{5}{3}-\beta}\,
[1-{\cal O}(\ell^{-\epsilon})]\, , \quad  \epsilon > 0\quad .
\label{eq7b}
\end{eqnarray}
Here $a'_{\ell ,1}$ denotes the position of the first maximum of the 
spherical Bessel function $j_\ell$ and $c_\beta$ denotes some finite
constant. 
We should point out that the latter
inequality (\ref{eq7b}) strongly overestimates the function 
$F_\ell$ by a constant value, which is 
even larger than the first (the largest) maximum of the function itself.
Although the integral over $t$ in Eq. (5) is convergent, one should be
careful when treating the high-energy ($t\rightarrow 1$) region, 
which can actually now generate unphysical spurious terms as a consequences
of the approximations performed above.  
We could restore or simulate
the asymtotics by introducing some appropriately chosen convergence factor,
e.g., $e^{-\mu t/(1-t)}$ with $\mu > 0$, if needed. 

Taking the absolute value of Eq. (\ref{eq6}) together with the approximations
(\ref{eq7a}) and (\ref{eq7b}) we can write
\begin{eqnarray}
\left|{\cal E}_L^a (\Lambda)\right| \leq 
\frac{2\alpha}{\pi}\,\sum_{\ell = L}^{\infty}\,
\left|\int_0^1\, dt\,\frac{t^{2+\beta}}{(1-t)^{1+\beta}}\right|\,
\frac{c_\beta}{2}\,
\left(\ell + \frac{1}{2}\right)^{-\frac{2}{3}-\beta}
\left|\langle a| r^\beta\,(V_c (r) - \overline{V}_c)|a\rangle\right|
\quad .
\label{eq8}
\end{eqnarray}
We observe the occurrence of Eulers' Beta function in the
expression above as a consequence of the estimates performed. The 
Beta function is defined via the integral (see formulae (6.2.1) and
(6.2.2) in Ref. \cite{aas71})
\begin{eqnarray}
B(z,w) := \int_0^1\, dt\,t^{z-1}\,(1-t)^{w-1} = 
\frac{\Gamma (z)\,\Gamma (w)}{\Gamma (z+w)}  .
\label{eq9a}
\end{eqnarray}
Note, that for certain values of the arguments $z$ and $w$ the Integration 
over $t$ has to be extended into the complex plane. Having in mind the 
analytical continuation, we can write:
\begin{eqnarray}
\left|\int_0^1\, dt\,\frac{t^{2+\beta}}{(1-t)^{1+\beta}}\right| = 
\int_0^1\, dt\,|B(3+\beta,-\beta)| = \int_0^1\, dt\,
\frac{\Gamma (3+\beta) \Gamma (1-\beta)}{\beta\,\Gamma (3)}
\quad .
\label{eq9b}
\end{eqnarray}
The analytical continuation of the Beta function is provided by the
reflection formula for the Gamma function (see (6.1.17) in Ref. \cite{aas71}):
$\Gamma (1-w) = -w\,\Gamma (-w) = \pi/\sin (\pi w)$ for 
$0 < \Re \{w\} < 1$.
The rigorous treatment of the integral (\ref{eq9a}), 
when evaluating it for the particular values $z = 3+\beta$ and $w = -\beta$, 
is provided by contour integration (see e.g. Ref. \cite{bat53}). 
It is performed along Pochhammer's 
closed contour ${\cal C}_P$ on the Riemann surface 
of the integrand $t^{z-1}(1-t)^{w-1}$ and relates the integral 
in Eq. (\ref{eq8}), respectively, the Beta function (\ref{eq9a}) for arbitrary 
arguments $z$ and $w$ to a product of Gamma functions according to
\begin{eqnarray}
e^{-\pi i\,(z+w)}\,\oint_{{\cal C}_P}\,\frac{dt\,t^{z-1}}{ (1-t)^{1-w}} &=&
 e^{-\pi i\,(z+w)}\,\left[1 - e^{2\pi i\,w} + e^{2\pi i\,(z+w)}
-e^{2\pi i\,z}\right]\,\int_0^1\,\frac{dt\,t^{z-1}}{ (1-t)^{1-w}}\nonumber\\
&=& -4 \sin (\pi z)\,\sin (\pi w)\,B(z,w) \quad .
\label{eq9d}
\end{eqnarray}
Thus, we are lead to the final expression for the finite majorante 
series 
\begin{eqnarray}
\sum_{\ell = L}^{\infty}\,
\left|f_\ell^a\left(\frac{t}{1-t}, \Lambda\right)\right| \leq
\frac{\Gamma (3+ \beta) \Gamma (1-\beta)\,c_\beta}
{4\beta}\,\left|\langle a| r^\beta\,
(V_c (r)- \overline{V}_c)|a\rangle \right|
\sum_{\ell = L}^{\infty}\, \left(\ell + \frac{1}{2}\right)^{-2/3-\beta}
\quad .
\label{eq10}
\end{eqnarray}
For $1/3 < \beta < 1$ the sum over $\ell$ is convergent and can be expressed
in terms of incomplete Zeta functions. 
Having derived a convergent 
majorante series the uniform convergence for the functional series
(\ref{eq3}) is proven. This allows indeed the interchange of the 
summation over $\ell$ with all the integrations involved. As 
result the generic correction term (\ref{eq1}) is equal to zero.
Thus, the spurious logarithmic terms derived by Persson {\em et al.} 
\cite{pss98} do not occur. 
This demonstrates the equivalence between the covariant 
renormalization and the non-covariant PWR approach when applied to the
problem under consideration. 

\subsection{Comment on the missing term}
We would like to point out that the evaluation of the correction term 
(\ref{eq1}) as it has been performed in Ref. \cite{pss98}
is incomplete and that the logarithmic
correction term Eq. (\ref{eqin}) appears as an artifact of the way
the expression has been evaluated. 
In Ref. ~\cite{pss98}, 
Eq. (\ref{eq1}) has been evaluated according to limiting
process:
\begin{eqnarray}
{\cal E}_1^a (\Lambda) + {\cal E}_2^a (\Lambda) &=&
 \frac{2\alpha}{\pi} \sum_{\ell=0}^{\infty} 
\lim_{K\rightarrow \infty} 
\left[ \int_0^{K \tilde{r}/r} dk\,f^a_\ell (k, \Lambda) + 
\int_{K\tilde{r}/r}^{K} dk\,f^a_\ell (k, \Lambda)\right] \quad , 
\label{eq11}
\end{eqnarray}
where $\tilde{r}$ is some average value of the coordinate $r$.
Following the arguments in Ref. \cite{pss98} the second term ${\cal E}_2^a$ 
vanishes in the limit $K\rightarrow \infty$, while the first part
${\cal E}_1^a$ generates the spurious term (\ref{eqin}).
In contrast to Ref. \cite{pss98} let us now take into account the third term,  
which is supposed to be zero when $K$ tends to infinity:
\begin{eqnarray}
{\cal E}_3^a (\Lambda) &=& \frac{2\alpha}{\pi} \sum_{\ell=0}^{\infty}\, 
\lim_{K\rightarrow \infty} 
\int_{K }^{\infty}\,f^a_\ell (k, \Lambda) 
\label{eq12}
\end{eqnarray}
Consider the sum of the second and third term, i.e.: 
\begin{eqnarray}
{\cal E}_2^a (\Lambda) + {\cal E}_3^a (\Lambda) &=&
\frac{2\alpha}{\pi} \sum_{\ell=0}^{\infty}\,
\lim_{K\rightarrow \infty} \lim_{\mu \rightarrow 0}
 \int_{K \tilde{r}/r}^{\infty} dk\,e^{-\mu k}\,f^a_\ell (k, \Lambda)
\label{eq13}
\end{eqnarray}
The regulator $e^{-\mu k}$ is
introduced for reasons of simplicity in order to guarantee a finite 
integral over $k$ at the upper integration limit, if the factor 
$j^2_\ell$ is absent. 
For finite values of the parameter $\mu$ this regularization will 
generate some large but $r$-independent constant, 
which, however, cancels in the matrix element. 
Similarly, one may include the regulator $e^{-\mu /k}$
in ${\cal E}_1^a$ to derive the same result (\ref{eqin}).

Now we can employ the same arguments that were used in Ref. \cite{pss98}
for calculating ${\cal E}_1^a$. The authors of \cite{pss98} argue that the
order between the integration over $r$ and the summation over $\ell$ can be
interchanged for the following reasons: a) the $r$-independent part of the 
$k$-integrand does not contribute to the matrix element 
$\langle a| \cdots |a\rangle$, and b) the convergence of the $k$-integral then 
does not depend on the Bessel function $j_\ell^2$.

The rigorous treatment of interchanging summations and integrations
requires the validity of Weierstrass' criterion as we have demonstrated 
in part A. 
Employing the arguments a) and b) above,
we assume that in Eq. (\ref{eq13})
the summation can be interchanged with all integrations 
and limits involved. Using the identity Eq. (\ref{eq2})
we arrive at
\begin{eqnarray}
{\cal E}_2^a (\Lambda) + {\cal E}_3^a (\Lambda)
&=& \frac{2\alpha}{\pi} \lim_{K\rightarrow \infty} \lim_{\mu \rightarrow 0}
\langle a| (V_c(r) -\overline{V}_c)  
\int_{K\tilde{r}\mu/r}^{\infty} 
\frac{d\xi\,e^{-\xi}} 
{\xi \sqrt{1+(\frac{\mu\Lambda}{\xi})^2}\left[
1+\sqrt{1+\left(\frac{\mu\Lambda}{\xi}\right)^2}\right]^2}|a\rangle
\nonumber \\ 
&=& \frac{\alpha}{2\pi} \lim_{K\rightarrow\infty}
\lim_{\mu \rightarrow 0}\, \langle a| (V_c(r) -\overline{V}_c) 
\left[E_1 \left(\frac{K\tilde{r}\mu}{r} \right)
+ {\cal O}\left(\frac{\mu^4 \Lambda K \tilde{r}}{r}\right)\right]
|a\rangle \quad.
\label{eq14}
\end{eqnarray}
To obtain this result we performed the limit $\mu\rightarrow 0$ in the 
integrand first. Expansion of the exponential integral for small arguments:
$E_1(\frac{K\tilde{r} \mu}{r} )
= \left(- \gamma -\ln\frac{K \tilde{r} \mu}{r} \right)
+{\cal O}\left(\frac{K \tilde{r} \mu}{r}\right)$ and 
taking the limit $\Lambda \to \infty$ of Eq. (\ref{eq14}) one ends up with
\begin{eqnarray}
{\cal E}_2^a + {\cal E}_3^a &=&  \frac{\alpha}{2\pi}\, 
\langle a| (V_c(r) -\overline{V}_c)\,\ln (r)|a\rangle\quad .
\label{eq15}
\end{eqnarray}
Thus, we obtain the logarithmic potential term as the remaining
cutoff-independent contribution.
It carries an opposite overall sign and will cancel in the total sum.
Moreover, we could introduce a unique regulator $e^{-\mu (1/k + k)}$ 
in expression (\ref{eq1}) and thus avoiding any (unnecessary) decomposition
of the $k$-integral and the introduction of $\tilde{r}$. 
Interchanging and performing the summation over $\ell$
at first one is left with 
\begin{eqnarray}
{\cal E}^a (\Lambda) 
= \frac{\alpha}{2\pi}\,
\lim_{\mu \rightarrow 0}\, \langle a| (V_c(r) -\overline{V}_c) 
2 \left\{K_0(2\mu) - \Lambda^2 K_2(2\mu) + \frac{15}{16} \Lambda^4
K_4 (2\mu) + \cdots \right\}|a\rangle  = 0 \quad .
\end{eqnarray}
All terms in the curly brackets vanish when the matrix element is
evaluated.
The calculation performed above indicates that the occurrence of the
spurious logarithmic term (and may be others as well) strongly depends 
on the analytical and on the numerical treatment of the high-energy
contribution to the SE.

\section{A numerical analysis on spurious terms}
In Ref. \cite{Yer} the occurance of spurious terms has been reported 
in connection with numerical evaluations
of the SE correction 
in one-electron ions with nuclear charge $Z$ in the presence
of additional Coulombic perturbation potentials. We will show, however, that
this is due to an improper treatment of the high-energy  contribution to the SE.
The total expression for the SE correction of a bound-electron state 
$|a\rangle$
can always be represented in the form (see  \cite{GLN} for
details):

\begin{eqnarray}
\Delta E^{{\rm SE}}_a = \sum_{n} \langle a n |\hat{\Sigma} |n a \rangle ,
\end{eqnarray}
where $\hat{\Sigma}$ denotes the nonlocal electron self-energy operator and the labels
$n$ run over the complete Dirac spectrum. 
The SE correction in the external field together with an additional
perturbation potential 
can be divided into three parts :
the wave-function correction, the vertex correction and the derivative
(or reference state) correction \cite{pss98}. We will concentrate here on the 
wave-function correction due to the self energy (SE, WF).
This correction can be obtained from the lowest order 
$\Delta E^{{\rm SE}}_a$ correction
by a replacement of the unperturbed wave function $|a\rangle$ by its first-order
perturbation theory correction

\begin{eqnarray*}
|a\rangle \rightarrow {\sum_{n}}' 
\frac{|n\rangle\langle n|V|a\rangle }{E_a -E_n} , 
\end{eqnarray*}
where $V$ denotes the perturbation potential. The prime indicates that the
term with $E_a = E_n$ is omitted from the sum. The energy
correction $\Delta E^{{\rm SE,WF}}_a$ can be written as

\begin{eqnarray}
\Delta E^{{\rm SE,WF}}_a = {\sum_{n,m}}' 
\left\{ \frac{\langle an |\hat{\Sigma} |nm \rangle
\langle m|V|a\rangle }{E_a-E_m} + 
\frac{\langle a|V|m\rangle \langle mn |\hat{\Sigma} |na \rangle}{
E_m-E_a} \right\}.
\end{eqnarray}
Within the B-spline approximation the complete Dirac spectrum
is represented by a purely  discrete one and 
terminates at some large number 
$N = 4 L (\nu + s -2)$, which is determined by the number of partial waves
$L$, the order $s$ of the B-spline functions
and the number of grid points $\nu$. 
We may restrict the summation over the energy of the
intermediate states in both Eqs. (19) and (20) by the condition
$|E_n| \le k \cdot mc^2$. The numerical results for the energy corrections as 
a function of the cutoff parameter
$k$ are shown in Table 1. What the lowest-order SE (Eq. (19)) is concerned 
its exact value in $U^{91+}$ ($Z$=92) 
is already obtained for $k=5$ within an accuracy of about 0.7\%.
Any further enlargement of the summation interval does not lead to any
improvement of the accuracy. We should note that the accuracy also depends on
the number of partial waves taken into account and 
on the number of grid points.
In our case we typically used $L$=6, and $\nu \simeq 140$
together with spline functions of order $s=9$. The quoted
accuracy is sufficient for our purposes since the 
contribution of the spurious term obtained in \cite{Yer} is
supposed to be much larger (about 20\%).

From Table 1 we conclude that the summation interval with $k=5$ for H-like $U^{+91}$
provides a sufficient approximation (about 2\% deviation) for 
$\Delta E^{{\rm SE,WF}}_a$.
However, we have found that any further
enlargement of this interval generates the spurious contribution 
as obtained in \cite{Yer}. From the
physical point of view the 20\% contribution that originates from the 
high energies larger then 5$mc^2$
(compared to the binding energy 0.3$mc^2$) is hardly understandable. Therefore, we conclude that the
occurance of the spurious contribution in the numerical calculations reported 
in Ref.\cite{Yer} has a similar
origin as the spurious term (1) derived analytically in 
Ref. \cite{pss98}, i.e. an improper
treatment of the high-energy contribution to the SE 
within the PWR approach.

\section{Calculation of QED corrections to the $g$-factor.}
Now we show that the conclusion drawn in Ref.\cite{pss98} concerning the
inapplicability of the PWR approach to SE calculations in 
external magnetic fields also 
does not hold strictly. For this purpose we employ a new approach to the
evaluation of the SE including an external magnetic field in the
Dirac equation from the beginning.

Accordingly,  the problem reduces again to the evaluation of the lowest-order SE 
(Eq. (19)). The Dirac equation with an external magnetic field is solved  
by means of an approach due to Chen and Goldman which we discribe 
briefly in Appendix B.

Within this approach the vacuum-polarization and self-energy corrections to the electron
$g$-factor in hydrogenlike heavy ions will be calculated. At first we calculate the
vacuum-polarization effect in order  to determine the values of the magnetic field
strength where this method remains stable. After this we turn 
to the calculation of the self-energy
correction  in the external magnetic field within the PWR method.

\subsection{Vacuum-polarization corrections to the electron 
$g$-factor of hydrogenlike heavy ions}

Vacuum-polarization (VP) corrections to the electron $g$-factor for H-like 
highly-charged (HCI) have been calculated within the 
the Uehling approximation \cite{uehl}.
For calculations of energy level of HCI 
this approximation is valid with an accuracy of about 10\% 
for all $Z$ values. 
Within the Uehling approximation the VP correction of a bound-electron
state $|a(B)\rangle$ including the external magnetic
field is determined by the matrix element:
\begin{equation}
\Delta E^{{\rm VP}}_a (B) = \langle a(B)|-\frac{Z}{r} S(r)|a(B)\rangle\, \equiv
\left(-\frac{Z}{r} S(r) \right)_{a(B) a(B)} ,
\end{equation}
where
\begin{equation}
S(r)=\frac{2\alpha}{3\pi} \int_1^{\infty} e^{-2rx /\alpha}\left(1+\frac{1}{x^2}
\right) \frac{\sqrt{x^2+1}}{x^2} \, dx
\end{equation}
and $\alpha$ denotes 
the fine structure constant (in atomic units $\alpha=1/c$). 
Then the correction to the $g$-factor results as
\begin{equation}
\delta g_a^{{\rm VP}} = \frac{2[\Delta E_a^{{\rm VP}}(B)-\Delta E_a^{{\rm VP}}(0)]}{\mu_B B},
\end{equation}
where $\mu_B$ is Bohr's magneton (in atomic units $\mu_B=\frac{1}{2}$).

The results of the calculations for the $1s$-ground state of H-like ions in comparison
with perturbation-theory results obtained
within the Uehling approximation \cite{Beier} are given in Table 2. An
agreement better than 0.01\% 
has been found for all $Z$ values from $5$ to $90$. In these
calculations values for the magnetic field strength 
$B=|\vec B|$ within the range of 
$1.0$ a.u. $\le B \le 100.0$ a.u. have been used.
The numerical results are stable within this range of $B$ values.

\subsection{Self-energy correction to the electron $g$-factor in hydrogenlike
heavy ions}

For the evaluation of the self-energy
correction in the external magnetic field we employ
the original PWR scheme developed in \cite{pls93,qag93} in
combination with the approach due to Chen and Goldman \cite{gold}
for solving the Dirac equation in cylindically-symmetric
external magnetic fields.

The formula for the SE correction in an external magnetic field for the state 
$a(B)$ reads

\begin{equation}
\Delta E_a^{{\rm SE}} (B) =
\frac{\alpha}{\pi} \sum_{n(B)} \left( \frac{1-\vec{\alpha}_1
\vec{\alpha}_2}{r_{12}} \int_0^{\infty} 
\frac{\sin(\kappa r_{12})\,d\kappa}{\kappa
-E_n(1-i0)+E_a} \right)_{a(B) n(B) n(B) a(B)} -\delta m_a (B),
\end{equation}
where the sum runs over the total Dirac spectrum,
$\vec{\alpha}$ denote the Dirac matrices, $r_{12}=|\vec{r}_1 - \vec{r}_2|$
and $\delta m_a$ abbreviates the counterterm. According to \cite{pls93,qag93} this
counter\-term follows from Eq. (24)
by replacing the summation over bound-states
$n(B)$ by a corresponding integration over free - electron Dirac states and by replacing
the bound-state
energy $E_n(B)$ in the denominator by the free-electron
energy $\epsilon$. The bound states $a(B)$ are expanded in free electron Dirac states
$|\vec{p},\epsilon>$ (where $\vec{p}$ is the electron momentum).
The correction to the $g$-factor that arises from the SE contribution reads
\begin{equation}
\delta g_a^{{\rm SE}} = \frac{2[\Delta E_a^{{\rm SE}}(B)-\Delta E_a^{{\rm SE}}(0)]}{\mu_B B}.
\end{equation}

Numerical results for the groundstate of H-like ions with different
$Z$ values are given in Table 3 in comparison with data 
obtained via perturbation theory
\cite{Beier} and from the $\alpha Z$-expansion (for low $Z$), respectively.
As for the VP corrections numerically stable results are provided 
for values of magnetic field strengths within the same range
$1.0$ a.u. $\le B \le 100.0$ a.u. 
For all $Z$ values
the deviations from the perturbation-theory results turn out to be smaller
than 0.1\%.

Thus, contrary to the statement made in \cite{pss98}, this provides numerical
evidence for the absence
of the spurious terms in calculations of
the SE correction in external magnetic fields
performed within the PWR approach
field at a level of accuracy better than 0.1\%.

We should note that the accuracy achieved in our approach will be not sufficient for obtaining
accurate values for bound-state QED corrections to 
electron $g$-factors. The reason  traces back to the fact that
bound-state QED corrections are obtained via subtraction of the free-electron QED
corrections from the values given in Tables 2 and 3. 
This leads to severe numerical cancellations that deminish
the accuracy of the net result significantly. However, it does not 
effect our
conclusions concerning the spurious terms.

\section{Conclusions}
Summarizing, we have provided evidence for
the absence of the "renormalization corrections"
for the particular case of the SE in external fields.
Contrary to the statements made in 
\cite{pss98}, \cite{Yer} we have found that 
via a proper treatment of the high-energy contribution
to the SE such corrections can be avoided numerically and analytically.
However, the proof presented here cannot exclude the 
possible existence of spurious
corrections in more complicated situations, e.g., in 
connection with  high-order QED corrections.
In the particular case of the "loop after loop"
second-order electron self-energy correction a discrepancy between
results obtained within the PWR
\cite{lal99} approach and from the covariant remormalization scheme
\cite{MS} and \cite{Yer1} has been reported.
This may indicate that this problem requires further investigation.

\section{Acknowledgements.}
Valuable discussions with S. Salomonson and A. Nefiodov are gratefully
acknowledged. I.G. is grateful to 
the Technical University of Dresden for the hospitality during his 
visit in 2000. 
The work of I.G. and L.L. has been 
supported by the RFBR grant $N^{o}$ 99-02-18526. 
G.S. and G.P.  acknowledge financial support from BMBF, DFG and GSI (Darmstadt).

\begin{appendix}

\section{The upper limit for Bessel functions.}
Let us consider at the beginning the real functions 
$F_\ell (x,\beta) = j_\ell^2(x)/x^\beta$ for values $ \beta > 0$ and
$\ell \geq L \gg 1$ over the interval $x\in [0, a_{\ell ,1}]$. 
We adopt the notation of Ref. \cite{aas71}. Let $a_{\ell ,1}$ and 
$a'_{\ell ,1}$ denote the location of the first zero and of first maximum 
of the spherical Bessel function $j_\ell (x)$, respectively. 
We employ the formulae (10.1.59) and (10.1.61) of 
Ref. \cite{aas71}:
\begin{eqnarray}
a_{\ell,1}^{'} &\approx& \left( \ell+ \frac{1}{2}\right )
\left[1 + 0.8086 \left( \ell+ \frac{1}{2}\right )^{-2/3}\right] 
\left( 1-{\cal O}(\ell^{-\epsilon} )
\frac{}{}  \right)\quad , \\
\label{eqa1}
j_\ell (a_{\ell ,1}^{'} )&\approx& 0.8458 \left(\ell+ \frac{1}{2}\right)^{-5/6}
\left( 1 - {\cal O}(\ell^{-\epsilon}) \frac{}{} \right)
\label{eqa2}
\end{eqnarray}
with $\epsilon > 0$.
Obviously, the zeros of the functions $F_\ell$ and $j_\ell$ coincide. For
any $\ell > 0$ and $\beta > 0$ the function
$F_\ell$ has exactly one maximum within the interval 
$x \in [0, a_{\ell ,1}]$ located at $x = b'_{\ell ,1}$, i.e.,  
$F'_\ell (b'_{\ell ,1},\beta) = 0$ and $F''_\ell (b'_{\ell ,1},\beta) < 0$. 
Derivatives with respect to $x$ are indicated by primes. 
For the first derivative $F'_\ell$ evaluated at $a'_{\ell ,1}$, we find
\begin{eqnarray}
F'_\ell (a'_{\ell ,1},\beta) = 
\left( \frac{2xj_\ell (x)j^{'}_\ell (x) - \beta\,j^2_\ell (x)}
{x^{\beta +1}}\right)_{\left|{x=a_{\ell ,1}^{'}}\right.} =
-\beta\,\frac{j^2_\ell (a_{\ell ,1}^{'})}{(a^{'}_{\ell ,1})^{\beta +1}} = 
-\beta \frac{F_\ell (a'_{\ell ,1},\beta)}{a'_{\ell ,1}} < 0 \quad .
\label{eqa3}
\end{eqnarray}
This implies: 1. $b'_{\ell ,1} < a'_{\ell ,1}$ and 2. $F_\ell$ decreases  
monotonically over the interval $b'_{\ell ,1} < x \leq a'_{\ell ,1}$.
Using the differential equation for the $j_\ell$ we derive for
the second derivative 
\begin{eqnarray}
F''_\ell (a'_{\ell ,1},\beta) &=& 
\left( \beta (\beta +1)\,\frac{j^2_\ell (x)}{x^{\beta +2}}
+2\, \frac{j_\ell (x) j''_\ell (x)}
{x^{\beta}}\right)_{\left|{x=a_{\ell ,1}^{'}}\right.} \nonumber \\
&=& \left\{ \frac{}{}\beta (\beta +1)-2 
\left[ \frac{}{}(a_{\ell ,1}^{'})^2- 
\frac{}{}\ell (\ell +1) \right]\right\} \,
\frac{j_\ell^2(a_{\ell ,1}^{'})}{(a_{\ell ,1}^{'})^{\beta +2}} < 0\quad ,
\label{eqa4}
\end{eqnarray}
which holds for all values $0 < \beta < L$, 
since $\ell \gg 1$. The fact that $F''_\ell (a'_{\ell ,1},\beta) < 0$ reveals
that $F_\ell$ has its point of inflection somewhere between
$a'_{\ell ,1}$ and $a_{\ell ,1}$, i.e. $F_\ell$ is a convex function 
over the interval $[b'_{\ell ,1}, a'_{\ell ,1}]$. 
Accordingly, we can estimate the value of $F_\ell$ at the position
$b'_{\ell ,1}$ of its first maximum, which finally provides an upper 
bound for the function $F_\ell$ throughout the whole range 
$x\in [0, \infty)$:
\begin{eqnarray}
F_\ell (b'_{\ell ,1},\beta) &\leq& F_\ell (a'_{\ell ,1},\beta) +  
|F'_\ell (a'_{\ell ,1},\beta)|\cdot |b'_{\ell ,1} - a'_{\ell ,1}|   
\leq (1 + \beta)\,F_\ell (a'_{\ell ,1},\beta) \quad .
\label{eqa5}
\end{eqnarray}
Here we used Eq. (\ref{eqa3}) and the fact that 
$|b'_{\ell ,1} - a'_{\ell ,1}| \leq a'_{\ell ,1}$. Finally we arrive at
\begin{eqnarray}
F_\ell (x,\beta) &\leq& 
(1+\beta) \,\frac{j_\ell^2(a_{\ell ,1}^{'})}{(a'_{\ell ,1})^\beta}
\leq c_\beta \left(\ell + \frac{1}{2}\right)^{-5/3-\beta}
\left[1 - {\cal O} (\ell^{-\epsilon})\right]\quad .
\label{eqa6}
\end{eqnarray}
What the proof is concerned we shall restrict to values $0 < \beta < 1$
only.

To provide  the function $F$ as required along the Pochhammer's contour we
need to extend the considerations above into the complex plane.
The function $F$ is analytic in the complex plane
$z=x+i\zeta$. A suitable region of the complex $t$-plane  (see Eq. (4))
which includes Pochhammer's contour is depicted in Fig. 1.

For small imaginary parts $0 < \varepsilon \ll 1$ all steps of the derivation
go through for complex arguments
$z=x+i\zeta$  with $|\zeta|\leq \varepsilon$. 
The Taylor expansion yields
\begin{eqnarray}
\left|F(z,\beta) \right|=\left|F(x,\beta) + i\zeta 
\left(\frac{\partial}{\partial\zeta}F(z,\beta)\right)\right|_{\zeta =0} 
\left. +  {\cal O} (\zeta^2)\right| \leq 
\left|F(x,\beta) \right|+\varepsilon
\left|F^{'}(x,\beta) \right|+  {\cal O} (\varepsilon^2) \quad,
\end{eqnarray}
where $0 < x < 1$. 
Now we might define ${{\rm sup}}_{(0<x<\infty)} |F^{'}(x,\beta)| \equiv
M<\infty$ together with
$\varepsilon=\left(\ell+1/2 \right)^{-5/3-\beta}$. 
In view of Eq. (A6) we can write

\begin{eqnarray}
 \left|F(z,\beta) \right| \leq (c_\beta+M)
 \left(\ell + \frac{1}{2}\right)^{-5/3-\beta} \left[1 - {\cal O} 
(\ell^{-\epsilon})\right] \quad .
\end{eqnarray}

\begin{figure}[b]
\centerline{\mbox{\epsfysize=6cm \epsffile{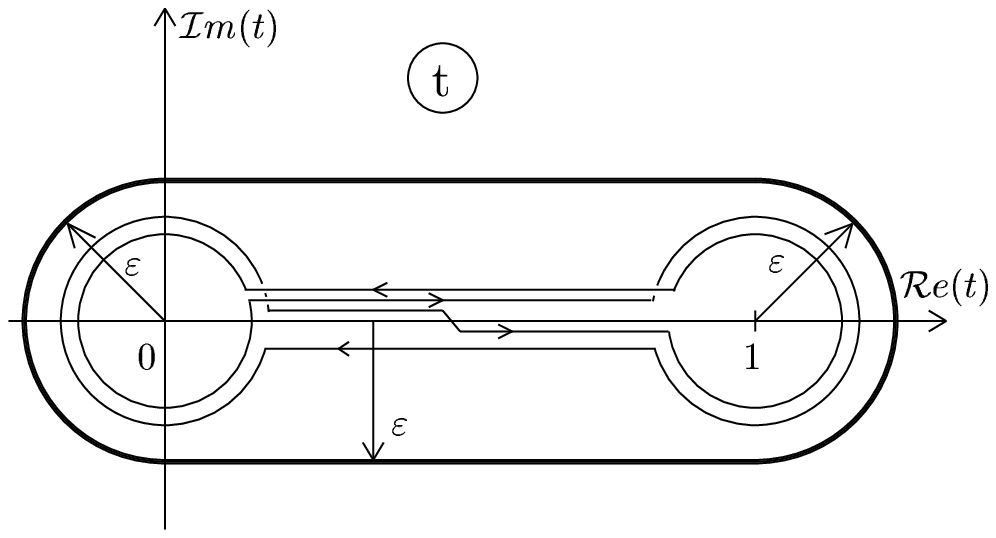}}}
\caption{The region of the complex $t$-plane for
which the approximation Eq. (A6) is valid.
It includes the Pochhammer's contour as depicted.}
\end{figure}

\section{Finite basis set solution of the Dirac equation for atomic
         electrons in external magnetic fields}

The Dirac equation for a bound electron in an 
additional external magnetic field reads
\begin{equation}
\hat{H} \Psi = E \Psi \, ,
\end{equation}
with
\begin{equation}
\hat{H}=\hat{H}_0+\hat{H}_m 
\end{equation}
and
\begin{eqnarray}
\hat{H}_0 &=& \vec{\alpha}\cdot\vec{p}+\beta m +V_{{\rm C}}(\vec{r}\,) \, , \\
\hat{H}_{{\rm m}} &=& \frac{1}{2} \vec{\alpha}\cdot [ \vec{B} \times \vec{r}\,] \, ,
\end{eqnarray}
where $\vec{\alpha}$, $\beta$ denote the Dirac matrices, 
$m$ is the electron mass, $\vec{B}$ is the magnetic field strength.
The magnetic field
is supposed to be directed along the $z$-axis: $\vec{B}=B\, \vec{e}_z$.
$V_{{\rm C}}(\vec{r}\,)$ is the Coulomb potential of a 
(point-like or extended) nucleus . 
In Eqs. (B1)--(B4) atomic units are used.

According to Ref. \cite{gold} 
the variational solution of the Dirac equation (B1) 
is obtained by means of trial functions
\begin{equation}
\Psi^{\mu}(\vec{r}\,)=\sum_{I=1}^{I_{{\rm max}}} \,
\sum_{\kappa}^{\kappa_{{\rm max}}}
a^{\kappa} \Psi_I^{\kappa \mu} (\vec{r}\,)\, .
\end{equation}

The electron wave functions $\Psi^{\mu}(\vec{r}\,)$
in the magnetic field possess cylindrical symmetry 
and can be expanded with respect to
a finite basis set of the functions $\Psi^{\kappa \mu}_{I}(\vec{r}\,)$ 
of spherical symmetry. The index $\kappa$ 
denotes the Dirac angular quantum number, $\mu$ corresponds to the total electron angular
momentum projection, $a^{\kappa} $ are the variational
coefficients, and $2\tilde{N}$ defines the number of the
basis-set functions.

The next step employs
the B-spline representation of the functions $\Psi^{\kappa \mu}_I(\vec{r}\,)$:
\begin{equation}
\Psi^{\kappa \mu}_I(\vec{r}\,)=\sum_{J=1}^{2N} b^{\kappa}_{IJ} \Phi_J^{\kappa
\mu} (\vec{r}\,)\, ,
\end{equation}
with
\begin{equation}
\Phi_n^{\kappa \mu}(\vec{r}\,)= \psi_n^{\kappa}\left(
\begin{array}{c}
i \frac{1}{r} \chi_{\kappa \mu} (\vec{r}/r) \\
0
\end{array} \right) \, ,
\end{equation}
\begin{equation}
\Phi_{N+n}^{\kappa \mu}(\vec{r}\,)= \psi_n^{\kappa}\left(
\begin{array}{c}
0 \\
-i \frac{1}{r} \chi_{-\kappa \mu} (\vec{r}/r) 
\end{array} \right),
\end{equation}
where $n=1,\dots,N$ and $\psi^{\kappa}_n$ denote
the B-spline representations of the finite basis set of radial functions
\cite{debor} and $\chi_{\kappa \mu} (\vec{r}/r)$ are the usual spherical
spinors.

The variational solution of the Dirac equation (B1) 
with the trial functions (B5) reduces 
to the diagonalization of the Hamiltonian (35) within the
finite basis set defined by Eq. (B5)--(B8). As the result 
one obtains the full
set of solutions of the Dirac equation for the atomic electron in an external 
magnetic field.

In particular, the matrix elements of the operator $\hat{H}_m$ with the wave functions
(B6) --  (B8) are given by:
\begin{equation}
\left\langle\Psi^{\kappa \mu}_n\left|
\hat{H}_{{\rm m}}\right|\Psi^{\kappa^{'}
\mu^{'}}_{N+n^{'}}\right\rangle
 = \frac{B}{2}\, r^{\kappa \kappa^{'}}_{nn^{'}}\, A_{\kappa \kappa^{'}}\, ,
\end{equation}
where
\begin{equation}
r^{\kappa \kappa^{'}}_{n n^{'}}=\int_0^{\infty} r ~ \psi_n^{\kappa} (r)
\psi^{\kappa^{'}}_{n^{'}} (r)\, dr\, ,
\end{equation}
\begin{equation}
A_{\kappa \kappa^{'}}=-i\int d\Omega ~ \langle\chi_{\kappa \mu}| [\vec{\sigma}\times
\vec{r}\,]_z | \chi_{-\kappa^{'} \mu^{'}}\rangle 
= \left\{ \begin{array}{l c r}
\frac{4\kappa \mu}{4\kappa^2 -1} & , {\rm for} & \kappa^{'}=\kappa \\
{\rm sgn}(\kappa) \frac{[(|\kappa|+1/2)^2 - \mu^2]^{1/2}}{2|\kappa|+1}  &,
{\rm for} & \kappa^{'}=-\kappa
\end{array} \right.
\end{equation}

It is convenient to rearrange the summations in 
Eq. (38) -- (39) in such a way 
that the summation over partial waves $L$ will be performened as the last one. 
In the B-spline approach
one starts to count the states from the lowest negative energy state ($J=1$). Then
the state with $J=N+1$ corresponds to the positive energy ground state  
$1s_{1/2}$.

Within the B-spline approach the H-like ion is included inside a spherical
box of radius $R_{{\rm box}}\sim \frac{50}{Z}\, a.u.$.
The number of grid points was $N_g=150$ and the order
of splines $k=8$. This corresponds to $2N=2(N_g+k-2)=314$ energy 
levels that represent approximately the Dirac spectrum. With this choice the
inaccuracy of the spline approximation for the $1s_{1/2}$ state compared
to the Chen and Goldman variational solution \cite{gold} becomes less than $10^{-8}$.
To test the accuracy of our approach we 
have calculated the Zeeman splitting of the
$1s_{1/2}$ state in the hydrogen atom for 
different field strengths $B$.

In Table 4 the results for the corresponding Zeeman splittings are compared
which have been evaluated 
within the approach Chen and Goldman and by means of perturbation theory.
The latter have been obtained employing the
standard formula \cite{landau}. The
comparison reveals that for field strengths $B$ up to $2 \cdot 10^2$ T
the deviation from the perturbation-theory results is about $10^{-8}$ 
while for a field strength of about $2 \cdot 10^4$ T
the deviation increase up to $10^{-3}$. 
The latter is due to the strong distortion of the atomic structure by the
magnetic field.

\end{appendix}

\begin{table}
\caption{$\Delta E_a^{{\rm SE}}$ and $\Delta E_a^{{\rm SE,WF}}$ for $U^{91+}$ as a function
of the energy cutoff $k$ of the
negative and the positive Dirac spectrum (see Eqs. (19--20)). All energy values in
a.u.. } 
\begin{tabular}{cccccc} \hline
& covariant & $k=5$ & $k=10 $ & $k=\infty$ & PWR \\
\hline
$\Delta E_a^{{\rm SE}}$ & 13.0714$^a$ & 12.9781 & 13.2190& 13.2114 &  \\  
$\Delta E_a^{{\rm SE,WF}}$ & 0.4626$^b$ & 0.4671 & 0.4914 & 0.5695& 
0.5598$^c$  \\ \hline
\end{tabular}
$^a$ Ref. \cite{Mohr}.\\
$^b$ Ref. \cite{Beier}. \\
$^c$ Ref. \cite{Yer}.
\end{table}

\begin{table}
\caption{VP corrections to the electron $g$-factor for the ground state of 
hydrogenlike ions.} 
\begin{tabular}{ccc} \hline
Z  &  $\delta g_{1s}^{{\rm VP}}$, this work & 
$\delta g_{1s}^{{\rm VP}}$, Ref. \cite{Beier} \\ \hline
1  & -3.0 $\cdot 10^{-11}$ &   \\
5  & -4.0 $\cdot 10^{-9}$ & -4.2$\cdot 10^{-9} $ \\
10 & -6.4$\cdot 10^{-8}$ & -6.37$\cdot 10^{-8}$ \\
20 & -9.37$\cdot 10^{-7}$& -9.41$\cdot 10^{-7}$ \\
50 & -3.32$\cdot 10^{-5}$& -3.32$\cdot 10^{-5}$ \\
90 & -4.3991$\cdot 10^{-4}$& -4.3995$\cdot 10^{-4}$ \\ \hline
\end{tabular} 
\end{table}
\begin{table}
\caption{SE corrections to the electron $g$-factor for the ground state of 
H-like ions. } 
\begin{tabular}{cccc} \hline
$Z$  &  $\delta g_{1s}^{{\rm SE}}$, this work & $\delta g_{1s}^{{\rm SE}}$, 
\cite{Beier} 
& $\delta g_{1s}^{{\rm SE}}$ $\alpha Z$-exp \cite{Beier} \\ \hline
1  & (2.31$\pm$0.01)$\cdot 10^{-3}$ & 2.322840$\cdot 10^{-3}$ &
2.32284$\cdot 10^{-3}$ \\
5 & (2.32$\pm$0.01)$\cdot 10^{-3}$ & 2.323389$\cdot 10^{-3}$ &
2.323388$\cdot 10^{-3}$ \\
10 & (2.321$\pm$0.009)$\cdot 10^{-3}$& 2.325472$\cdot 10^{-3}$ &
2.3249028$\cdot 10^{-3}$ \\
20 & (2.332$\pm$0.007)$\cdot 10^{-3}$& 2.33692$\cdot 10^{-3}$ & \\
50 & (2.469$\pm$0.012)$\cdot 10^{-3}$& 2.47162$\cdot 10^{-3}$ & \\
90 & (2.997$\pm$0.011)$\cdot 10^{-3}$& 3.04516$\cdot 10^{-3}$ & \\ \hline
\end{tabular} 
\end{table}

\begin{table}
\caption{Results for the Zeeman splitting \protect$\Delta E_{{\rm Z}}\protect$ 
for the ground state in 
neutral hydrogen obtained in the Chen and Goldman approach (CGA) 
\protect\cite{gold} and in 
perturbation-theory (PT). All values are given in atomic
units (a.u). For the magnetic field strength: 1 a.u. = 
\protect$2.35 \cdot 10^5\protect$ T.}
\begin{tabular}{ccc} \hline
$B$        &$\Delta E_{{\rm Z}}$ (CGA) &$\Delta E_{{\rm Z}}$ (PT) \\ \hline
0        & 0.5000066566    & 0.5000066566 \\
$10^{-5}$ & 0.5000116565    & 0.5000116565 \\
$10^{-3}$ & 0.500506398     & 0.5005066477 \\
$10^{-1}$ & 0.5497433       & 0.5500057691 \\
1        & 0.872133        & 0.9999977813 \\ \hline
\end{tabular}
\end{table}

\end{document}